\documentclass[letter,twocolumn]{IEEEtran}
\usepackage{cite}
\usepackage{amsmath,amssymb,amsfonts}

\usepackage{algorithmic}
\usepackage{graphicx}
\usepackage{tabularx}
\usepackage[table,usenames]{xcolor} 
\usepackage{array}
\usepackage{mathtools, cuted}
\usepackage{relsize}
\graphicspath{ {images/} }
\usepackage{textcomp}
\def\BibTeX{{\rm B\kern-.05em{\sc i\kern-.025em b}\kern-.08em
T\kern-.1667em\lower.7ex\hbox{E}\kern-.125emX}}
\usepackage{verbatim}
\usepackage[overload]{textcase}
\makeatletter

\newcommand{\Rmnum}[1]{\expandafter\@slowromancap\romannumeral #1@}
\newcommand{\comm}[1]{}
\usepackage{multicol}
 \usepackage{bbm}
\usepackage{amsthm}
\usepackage[hyphens]{url}
\usepackage[hidelinks, bookmarks=false, final=false]{hyperref}
\usepackage{booktabs}
\usepackage{caption} 

\DeclareRobustCommand{\element}[1]{\@element#1\@nil}
\def\@element#1#2\@nil{%
  #1%
  \if\relax#2\relax\else\MakeLowercase{#2}\fi}
\makeatother

\definecolor{lightblue}{rgb}{0.93,0.95,1.0}

\begin{document}

\title{Queuing with Deterministic Service Times and No Waiting Lines in Machine Type Communications}
% \title{On Multi-Server Queues with Degenerate Service Time Distributions and No Waiting Lines (G/D/n/n)}

\author{Ren\'{e}~Brandborg~S{\o}rensen,~\IEEEmembership{Student~Member,~IEEE, }Jimmy~J.~Nielsen,~\IEEEmembership{Member,~IEEE} 
and Petar~Popovski,~\IEEEmembership{Fellow,~IEEE} 
\thanks{René B. Sørensen, Jimmy J. Nielsen and Petar Popovski are with the Connectivity section, Electronic Systems, Aalborg University, Denmark,
 e-mail: (rbs, jjn, petarp)@es.aau.dk.}
\thanks{Manuscript submitted \today; revised -}}

\markboth{Wireless Communication Letters,~Vol.~X, No.~X, Month~Year}%
{S{\o}rensen \MakeLowercase{: Queuing with Deterministic Service Times and No Waiting Lines in Machine Type Communications}}

\maketitle

\begin{abstract}
The growth of Machine-Type Communication (MTC) increases the relevance of queuing scenarios with deterministic service times. In this letter, we present a model for queues without waiting lines and with degenerate service time distributions and show how the framework is extendable to model general service time distributions. Simple bounds and a close approximation of the blocking probability are derived and the results are shown to hold for simulated queues with Markovian and degenerate arrival processes.
\end{abstract}


% \fi
\IEEEpeerreviewmaketitle

\section{Introduction}
% \IEEEPARstart{Q}{}ueues are useful modelling tools in engineering. 
% Naturally, queues that describe common phenomena are useful. Many natural processes found in engineering are Poisson processes; the number of customers arriving at a bank, the number of customers being served in said bank, the number of cars passing a bridge or intersection, the number of people dialing a friend for a short call and the number of people ending said call.
% \IEEEPARstart{Q}{}ueues are useful for models in engineering. In particular, 
\IEEEPARstart{A}{}number of emerging use cases in Machine Type Communication (MTC) \cite{8401624} require queueing models that are significantly different from traditional models used in teletraffic theory. A distinctive feature of many MTC applications is that machines are likely to perform different tasks within an almost-deterministic time. This brings relevance to queues with degenerate service times and non-Markovian, in particular periodical, arrival times. While networks in general have become packet-switched, relying heavily on buffering, there are still switching operations within communications that require immediate service or service within a very short time. Let us, for example, take the case of LoRa, where messages are modulated with one of seven different spreading factors. In LoRaWAN six spreading factors are used to create six quasi-orthogonal sub-channels in each channel within the network. LoRa gateway transceiver chipsets are capable of detecting preambles for every spreading factor simultaneously on multiple channels, but only a finite amount of demodulation paths are available \cite{whencollide,lastPaper}. So messages in excess of the available demodulation paths are lost.
Other general examples in telecommunications include: service of critical real-time interrupts \cite{896384}, scheduling of immediate resources in FDMA networks and packet demodulation in FDMA networks.
% However, queues with a constant, deterministic service rate are common in IT and communication engineering. In particular, human-to-machine and machine-to-machine communications incite a shift from the random service times of humans, to the predefined, constant, deterministic service times of machines.

% \subsection{Related work}
We use Kendall's notation \cite{kendall1953}, noting that \texttildelow/\texttildelow/n/n refers to queues of finite capacity equivalent to the number of servers. The steady state solution for the M/D/n/n queue is well known as derived by A. K. Erlang. This solution was later shown to be valid for M/G/n/n queues \cite{onErlang}. %and extensively used in modelling teletraffic.

In this paper, a framework for modelling G/D/n/n queues is presented and a close approximation of and bounds on the blocking probability are derived. It is also shown how the framework is applicable to general service times distributions, that is, G/G/n/n queues.
The approximation is shown to comply with simulation for Markovian and Degenerate arrival distributions. In the Markovian case, this entails that it also fits well with well-known exact solutions for M/G/n/n~queues.
%\textcolor{blue}{check once more if there are other relevant works}

% is given and it is shown to be equivalent to previous results for the Erlang B (M/G/n/n)~queue.

%Furthermore, the M/D/n/n queue is explore for advanced engineering applications, specifically jobs with heterogeneous service times and servers with heterogeneous service times.
% \pp{The presented model and investigation of the G/D/n/n queue is applicable more generally, beyond the motivating scenarios with MTC.}
The presented model and investigation of the G/D/n/n queue is applicable more generally, beyond the motivating scenarios with MTC.

\section{System model}
% \pp{Do it with a device and a Base Station rather than Alice and Bob. I would start it by saying - let us consder a typical IoT scenario based on LoRa where .... This is modelled by ...} 
% Let Alice and Bob exchange messages via a multi-channel messaging system. Once a message is generated Bob or Alice immediately puts the message on a free channel, which is occupied for the message duration. 
% Consider a MTC device transmitting messages to a base station via a multi-channel messaging system. Once a device generates a message, \pp{it is immediately sent over a free channel, which is then occupied for the message duration.}
% Messages generated when no channels are available are discarded. A concrete scenario in MTC is that of reservation of demodulation paths in LoRa receivers \cite{whencollide,lastPaper}. Here, receivers are listening to and detecting preambles on multiple channels, but the number of demodulation paths available to receive orthogonal transmissions is limited, thus transmissions which happen to find all demodulation paths occupied are dropped. 

We will develop a general analytical model for queues with deterministic service times, but we will treat the concrete problem of reservations of demodulation paths in LoRaWAN gateways as mentioned in the introduction.
SX1301 is a chipset meant for usage in LoRaWAN gateways. This chipset is capable of demodulating up to 8 frames in parallel \cite{sx1301}.

\subsection{Arrival process}
Let the number of messages transmitted within a fixed time $T$ be denoted $k$. The probability of at-least $k_0$ transmissions within a time step of $\tau$ is  %$B_k(\lambda,\tau)$. %This probability is useful for describing the queue since each server can serve one customer within $\tau$. 
%This probability is given by \eqref{eq:p_prob1}. 
%
\begin{align}
    B_0(\lambda,\tau) = \Pr(k \geq 0 | \sum_{x\in\{\}} t_x \leq \tau, \lambda&) \mathrel{\mathop:}= 1 \;,\label{eq:p_prob1}
\\  B_k(\lambda,\tau) = \Pr(k \geq k_0 | \sum_{x=1}^{k_0} t_x\leq \tau, \lambda&) = \iint_D\left(f_{1,2,...,p}\right)dD \;,\nonumber
\\    D \in &\{t_1,...,t_x|\sum_{x=1}^{k_0} t_x\leq \tau\} \;.\nonumber
\end{align}

Then we can find the probability of transmitting exactly $k$ messages in a period $\tau$ by %$A_k(\lambda,\tau)$
\begin{align} \label{eq:A_k1}
A_k(\lambda,\tau) &= \Pr(k = p| \sum_{x=1}^p t_x\leq \tau, \lambda) \;,
\\ = \Pr(k &\geq p | \sum_{x=1}^p t_x\leq \tau, \lambda)-\Pr(k \geq p+1 | \sum_{x=1}^{p+1} t_x\leq \tau, \lambda)\;, \nonumber
\\ &= B_k(\lambda,\tau)-B_{k+1}(\lambda,\tau)\;, \quad\text{for} \quad 0\leq k\leq \infty \;.\nonumber
% \\= \text{Pois}(&p,\lambda\tau)
\end{align}

% \begin{figure} [tb]
%     \centering %[width=1\textwidth]
%     \includegraphics[width=1\columnwidth,height = 140pt]{figures/scenario.eps}
%     \caption{A representation of the M/D/n/n queue.}
%     \label{fig:scenario}
% \end{figure}
%\subsection{Intermittency in the arrival process}
The set of received messages is a subset of the set of transmitted messages due to outage caused by for example poor channel conditions, noise or interference. Let $p_o$ be the outage probability for a transmission not to be received. 
Then the probability for the number of received messages can be found by transforming the probability of the number of transmissions as
% Let an arrival process generate prospects. Prospects become costumers and enter the queue with probability $1-p_o$ \pp{PP: What is the interpretation of this probability? Is it due to channel errors or ...? In addition, this confirms my point from above, that we should use $k_0$ and reserve $p$ or any indexed version of it for probability. Relate this to LoRa demodulation paths.}. The arrival count probability for the customer arrival process can then be found by transforming the arrival count probability of the prospect arrival process as
\begin{align}
    A^\diamond_k(\lambda,\tau) &= \sum_{x=k}^\infty(A_k(\lambda,\tau)(1-p_o)^kp_o^{x-k}{k \choose x}) \;.\nonumber
\end{align}

\begin{figure}
    \centering
    \includegraphics[width=\columnwidth]{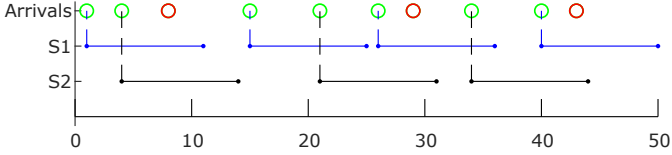}
    \caption{Example of arrivals, service times and blocking in a G/D/2/2 queue. Messages occupy servers for a fixed service time, $\tau$. Messages who arrive to find all servers occupied are blocked. }
    \label{fig:exQ}
\end{figure}

\subsection{Queue behaviour}
%We shall model the queue based on the perspective of a random, observant customer who arrives at a random time $t_{i-1}$. %\pp{How is this related to the motivating MTC scenario? What is a ``customer'' in this case and when it is blocked?}.
We denote the number of transmissions being demodulated at $t_{i-1}$ by $K_{i-1}$ and the number transmissions taking up demodulation paths after $t_{i-1} + \tau$ by $K_i$. Denote new arrivals in the queue by $K^\text{A}_i$, demodulated transmissions by $K^\text{D}_i$ and the number of blocked transmissions by $K^\text{B}_i$. transmissions are blocked when they arrive to find all demodulation paths unavailable as depicted in Fig~\ref{fig:exQ}.
Then we have %\PP{PP: Can you have a figure that illustrates this?}
\begin{align}
    K_i &= K_{i-1} + K^\text{A}_i - K^\text{D}_i - K^\text{B}_i\;,
    \\&&0\leq K_i\leq n \;.\nonumber
\end{align}

\begin{align}
    K^\text{B}_i &= {\max(K_{i-1}+Z_i-n,0)} \;,
    \\ \text{where}\quad Z_i &= K^\text{A}_i - K^\text{D}_i \;,\nonumber
    \\ \text{so}\quad f_{Z_i} &= f_{K^\text{A}_i}*\hat{f}_{K^\text{D}_i} \;,\nonumber
    \\ \text{where}\quad \hat{f}_{K^\text{D}_i}[k] &= f_{K^\text{D}_i}[-k] \;,\nonumber
   \\ \text{and}\quad f_{K^\text{A}_i}[k] &= A^\diamond_k \;.\nonumber
\end{align}
All messages that arrived within the prior period and weren't blocked are served, so
\begin{align}
K^\text{D}_i = \min(K^A_{i-1},n)
% f_{K^\text{D}_i}[k] = f_{K_{i-1}}[k]\;.
\end{align}
%\pp{For example, here you can relate the equations above to a LoRa or FDMA scneario as it stays too abstract. }

\section{Analysis}
\subsection{Bounds on the blocking probability}
The blocking probability is defined as
\begin{align} \label{eq:pb}
    P_{b} = \dfrac{\mathrm{E}[K^\text{B}_i]}{\mathrm{E}[K_i]} = \dfrac{\mathrm{E}[K^\text{B}_i]}{\mathrm{E}[K^\text{B}_i] + n}\;,
\end{align}
where the substitution $\mathrm{E}[K_i] = \mathrm{E}[K^\text{B}_i] + n$ in the denominator in \eqref{eq:pb} is valid, because the blocking probability is zero until the amount of messages in queue is larger than the total number of demodulation paths $n$. %, that is $K_i = K^\text{B}_i + n$ for $K^\text{B}_i > 0$ and $P_b = 0$ for $K_i\leq n$. 
This also means that
\begin{align}
f_{K^\text{D}_i}[k] &= 
    \begin{cases}
    1\;, & \text{for } k = n\;,\\
    0\;, & \text{otherwise}\;.
    \end{cases}    
\end{align}

Assume that there's no spill-over between messages in the observed periods $i-1$ and $i$, then $K_{i-1} = 0$ and we obtain a lower bound on the blocking probability.
\begin{align}
    P_{b_\mathrm{lower}} &= \dfrac{\sum\limits_{k=1}^\infty f_{K^\text{B}_i|K_i = n}[k]\cdot k}{\sum\limits_{k=1}^\infty f_{K^\text{B}_i|K_i = n}[k]\cdot k+n}\;,\nonumber
    \\&= \dfrac{\sum\limits_{k=n+1}^\infty(A_k(\lambda,\tau) \cdot (k-n))}{\sum\limits_{k=n+1}^\infty(A_k(\lambda,\tau) \cdot (k-n))+n} \;.\label{eq:pblow}
\end{align}

% Note that for $n=1$ there is no customer spill-over since they are served in exactly $\tau$, so the blocking rate is exactly equal to the upper bound \eqref{eq:pbhigh}.

In the same manner, we can assume that there's complete spill-over between messages in period $i-1$ and $i$, then $K_{i-1} = n$ and we obtain an upper bound on the blocking probability. %\pp{Explain this assumption in terms of LoRA?}
\begin{align}
    P_{b_\mathrm{upper}} &= \dfrac{\sum\limits_{k=1}^\infty f_{K^\text{B}_i|K_i = 0}[k]\cdot k}{\sum\limits_{k=1}^\infty f_{K^\text{B}_i|K_i = 0}[k]\cdot k+n}\;,\nonumber
    \\&= \dfrac{\sum\limits_{k=n}^\infty(A_k(\lambda,\tau) \cdot k)}{\sum\limits_{k=n}^\infty(A_k(\lambda,\tau) \cdot k)+n} \;.\label{eq:pbhigh}
\end{align}

\subsection{Blocking probability}
To describe the exact blocking probability we need to describe the spill-over between observation period $i-1$ and $i$, $K_{i-1}$. We let $X_i = K_{i-1}+Z_i$ so that
\begin{align}
    f_{X_i}[x_i] &= \sum_{k_{i-1}={0}}^n f_{K_{i-1},Z_i}[k_{i-1},x_i-k_{i-1}]\;,
    \\f_{K_{i-1},Z_i}[k_{i-1},z_i]  &= 
 \\f_{Z_i}[Z_i = &z|K_{i-1} = k_{i-1}]\cdot f_{K_{i-1}}[K_{i-1}=k_{i-1}] \;.\nonumber
\end{align}

Then we can describe the probability of blocking $k$ transmissions as
\begin{align}
    f_{K^\text{B}_i}[k] = 
        \begin{cases}
        \sum\limits_{x=-\infty}^nf_{X_i}[x]\;,  &\text{for } k = 0\;, \\
        f_{X_i}[k+n]\;,  &\text{for } k \geq 1\;, \\
        0\;, &\text{otherwise }\;.
        \end{cases}  
\end{align}

Then we can use a prior for $f'_{K_{i-1}}$ to approximate $f'_{K^\text{D}_{i}}$, $f'_{Z_i}$ and $f_{X_i} \approx \dfrac{f_{X_i}|f'_{K_{i-1}}}{\sum(f_{X_i}|f'_{K_{i-1}})}$ and then approximate the blocking probability, $P_b$, using \eqref{eq:pb}. 

In case a prior is not evident for an arrival process G, then we may use
\begin{align}
    f_{K_ {i-1}}[k] = \label{eq:generalprior}
        \begin{cases}
        \sum\limits_{x=-\infty}^0 f_{K^\text{A}_i}[X_i = x]\;, &\text{for } k = 0 \;,\\
        f_{X_i}[K^\text{A}_i = k]\;, &\text{for } 0 < k < n    \;,\\
        \sum\limits_{x=n}^\infty f_{K^\text{A}_i}[X_i = x]\;, &\text{for } k = n \;,\\
        0\;, &\text{otherwise }\;.
        \end{cases}    
\end{align}

% The time spent

\subsection{Server state probability and server utilization} \label{sec:timing}
% \textcolor{blue}{more on this?}
% The posterior probability of having $K_i$ users in queue at the end of the observation period is 
% \begin{align}
%     f_{K_i}|f_{K_{i-1}}[k] = 
%         \begin{cases}
%         \sum\limits_{x=-\infty}^0 f_{X_i}[X_i = x] &\text{for } k = 0 \\
%         f_{X_i}[X_i = k] &\text{for } 0 < k < n    \\
%         \sum\limits_{x=n}^\infty f_{X_i}[X_i = x] &\text{for } k = n \\
%         0 &\text{otherwise }
%         \end{cases}    
% \end{align}
The server utilization can be found by considering the timing within the queue. Consider the case of $n=1$, then the time spent without messages in queue can be described as the time between completion of service of one message till the arrival of the next. Hence:
\begin{align}
    T_0 &= \sum_{k=0}^\infty C_k \label{eq:T0}\;\quad\text{and}\quad T_1 = \tau \;,
\end{align}   
where $C_k$ is the mean time spent without a message in queue if the k'th message is the first one received after the demodulation of another message finishes,
\begin{align}   
    % C_0 &= \int_{t=0}^\infty \Pr(t_1 = t)\cdot( t-\tau)\; dt \nonumber
    C_k &= \int_{t=\tau}^\infty\Pr(\sum_{x=1}^{k+1}t_x = t|\sum_{x=1}^{k}t_x \leq \tau)\cdot (t-\tau)\; dt \label{eq:Ck}\;,
    \\ &\quad\text{for}\quad k\geq0 \nonumber\;.
\end{align}
The complexity of describing the timing in this way increases greatly as $n$ increases.
The state ratio for state $y$ is given by
\begin{align}
    q_y = \dfrac{T_y}{\sum T_y}\;.
\end{align}
Based on the state ratios we can compute the average number of messages being served as \eqref{eq:L} and the server utilization as \eqref{eq:eta}.
\begin{align}
    L &= \sum_{y=0}^nq_y\cdot y\;. \label{eq:L}
    \\ \eta &= \dfrac{L}{n}\;. \label{eq:eta}
    \\ \zeta &= \eta(1-P_b)\;, \label{eq:zeta}
\end{align}
where $\zeta$ is the non-blocking server utilization, which is the probability that the demodulation path is being used and message demodulation is not being blocked.

% \textcolor{blue}{Calculate T0 for Markov and Degenerate}

% \subsection{Waiting Lines}
% \textcolor{blue}{G/D/n/n+N} 

% Let $0\leq~K_i\leq~n+N$ then 
% \begin{align}
%     f_{K^\text{D}_i}[k] = 
%         \begin{cases}
%         f_{K_i}[k]  &\text{for } 0\leq k < n \\
%         \sum\limits_{k=n}^{n+N}f_{K_i}[k]  &\text{for } k =n \\
%         0 &\text{otherwise }
%         \end{cases}  
% \end{align}
% \begin{align}
%     f_{K^\text{B}_i}[k] = 
%         \begin{cases}
%         \sum\limits_{x=-\infty}^{n+N}f_{X_i}[x]  &\text{for } k = 0 \\
%         f_{X_i}[k+n+N]  &\text{for } k \geq 1 \\
%         0 &\text{otherwise }
%         \end{cases}  
% \end{align}

\subsection{Jobs with non-homogeneous service times}
We divide messages into classes $C_1$ through $C_m$ corresponding to the different spreading factors in LoRaWAN. The service time of class $x$ is given as $\tau_x~\in~\boldsymbol{\tau}~=~\{\tau_1,\tau_2,...,\tau_m\}$ for corresponding mean arrival rates of $\lambda_x~\in~\boldsymbol{\lambda}~=~\{\lambda_1,\lambda_2,...,\lambda_m\}$. %We denote the set of arrival rates for all customer classes as $C_\rho = \{(\lambda_1,\tau_1),(\lambda_2,\tau_2),...,(\lambda_m,\tau_m)\}$

Denoting the number of messages from class $y$ by $k_y$, the probability of $k_y$ messages arriving within the service period, $\tau_y$, is $A_k(\lambda_y,\tau_y)$ and the arrival distribution $f_{K^\text{A}_i}$, we have
\begin{align} \label{eq:ggnn}
f_{K^\text{A}_i}(\boldsymbol{\lambda},\boldsymbol{\tau}) &= \sum_{y=1}^mf_{K^\text{A}_i}(\lambda_y,\tau_y)\dfrac{\lambda_y}{\sum_{x=1}^m\lambda_x} \;.
\end{align}

Any service time distribution can be represented by binning with infinitesimally small bins, so that we obtain two infinite sets for $\boldsymbol{\lambda}$ and $\boldsymbol{\tau}$, where $\lambda_x~=~p_x~\cdot~\lambda$ and $p_x~=~f_{\text{ServiceDist}}(\tau_x)$. Practically, the service time distribution can be accurately approximated by binning with an appropriately small finite bin size. % as in Fig. \ref{fig:pdftau}, so that we obtain two infinite sets for $\boldsymbol{\lambda}$ and $\boldsymbol{\tau}$, where $\lambda_x~=~p_x~\cdot~\lambda$ and $p_x~=~f_{\text{ServiceDist}}(\tau_x)$. 
In this way we may be able to represent any G/G/n/n queue in the described framework by binning the service time distribution.

\section{Degenerate and Markovian arrival processes}
In this section the transmission count probability, $A_k$, is derived for degenerate and Markovian arrival processes by solving \eqref{eq:A_k1}. The mean arrival-rate, $\lambda'$, and priors, $f'_{K_{i-1}}$, are also discussed for each arrival process. 

\subsection{Degenerate inter-arrival distribution}
Let transmissions occur with inter-arrival times $t_x$ that are distributed according to a degenerate distribution; $f_{t}(t)~=~p_x$ for $t=t_x$ where $t_x\in\{t_1,t_2,...,t_y\}$ for a corresponding set $p_x\in\{p_1,p_2,...,p_y\}$ where $\sum_{x=1}^yp_x=1$.
Solving \eqref{eq:p_prob1} and \eqref{eq:A_k1} for the D/D/n/n queue we obtain %that the arrival count is distributed according to Eq.~\eqref{eq:A_kD}.

\begin{align} %\label{eq:A_kD}
    B_0(\lambda,\tau) &= \sum_Dp_x
    \;,\quad\text{for}\quad D={\{x|t_{x} \geq \tau\}} \nonumber \;,
    \\B_1(\lambda,\tau) &=  \sum_Dp_x \;,\quad\text{for}\quad D={\{x|t_{x} \leq \tau\}} \nonumber\;,
    \\B_2(\lambda,\tau) &=  \sum_D\prod_{i=1}^2p_{x_i} \;,\;\text{for}\; D={\{x_1,x_2|\sum_{y=1}^2t_{x_y} \leq \tau}\} \nonumber\;,
    \\B_k(\lambda,\tau) &=  \sum_D\prod_{i=1}^kp_{x_i} \;,\;\text{for}\; D={\{x_1,x_2,...,x_k|\sum_{y=1}^kt_{x_y} \leq \tau}\} \nonumber\;,
    \\A_k(\lambda,\tau) &=  B_k(\lambda,\tau)-B_{k+1}(\lambda,\tau)\;. \label{eq:A_kD}
\end{align}
The prior of \eqref{eq:generalprior} gives good approximations for $n>1$. When n=1, it is clear that $K_{i-1}= 0$ holds, so our prior should be $f_{K_{i-1}}[k] = 1 \;\text{for}\; k = 0$ where $f_{K_{i-1}}[k] = 0. \;\text{for}\; k \neq 0$ 
The mean arrival-rate is $\lambda' = \dfrac{1}{\sum_{x=1}^y\dfrac{p_x}{t_x}}$. $C_k$ can be found to be
\begin{align}
    C_k(\lambda,\tau) &=  \sum_D(\prod_{i=1}^{k+1}p_{x_i} \cdot (\sum_{i=1}^{k+1}t_{x_i}-\tau))\;,
    \\&\text{for}\; D={\{x_1,x_2,...,x_{k+1}|\sum_{y=1}^kt_{x_y} \leq \tau, \sum_{y=1}^{k+1}t_{x_y} > \tau}\} \;,\nonumber
\end{align}

\subsection{Exponential inter-arrival distribution}
Let transmissions occur with inter-arrival times $t_x$ that are distributed according to an exponential distribution; $f_{t}(t)~=~\lambda\exp(-\lambda t)$ for $t\geq0$. Solving \eqref{eq:A_k1} for this arrival process we obtain %that the arrival count is Poisson distributed as illustrated in Eq.~\eqref{eq:A_kM}.% as in \cite{MD1c}.
\begin{align} \label{eq:A_kM}
    % \Pr(A_1) &= \Pr(t_1 \leq \tau)-\Pr(t_1+t_2 \leq \tau) \nonumber
    % \\&= \lambda\tau
    A_0(\lambda,\tau) &= \exp({-\lambda\tau})  \nonumber\;, \\A_1(\lambda,\tau) &= \lambda\tau\cdot \exp({-\lambda\tau}) \nonumber \;,
    \\ A_2(\lambda,\tau) &= \dfrac{(\lambda\tau)^2}{2}\cdot \exp({-\lambda\tau}) \nonumber  \;,
    \\A_3(\lambda,\tau) &= \dfrac{(\lambda\tau)^3}{6}\cdot \exp({-\lambda\tau})  \nonumber \;,
    %\\ &\Pr(A_4(\tau)) = \dfrac{(\lambda\tau)^4}{24}\cdot \exp({-\lambda\tau}) \;,\; \Pr(A_5(\tau)) = \dfrac{(\lambda\tau)^5}{120}\cdot \exp({-\lambda\tau})  \nonumber
    %\\ &\Pr(A_6(\tau)) = \dfrac{(\lambda\tau)^6}{720}\cdot \exp({-\lambda\tau})  \quad,\quad \nonumber
    \\A_k(\lambda,\tau) &= \dfrac{(\lambda\tau)^k}{k!}\cdot \exp({-\lambda\tau}) \;.
\end{align}
The mean arrival-rate is $\lambda$. We shall use the prior in \eqref{eq:generalprior} when approximating $P_b$. $C_k$ can be found to be
\begin{align}
C_k(\lambda,\tau) = \dfrac{\exp{(-\lambda\tau)}\lambda^{k-1}\tau^k}{k!}\;.
\end{align}

\section{Results}
In this section we present results for the accuracy of this framework for modelling M/D/n/n and D/D/n/n queues for a fixed service time and a set of heterogeneous service times. Then we discuss the impact of the results on the example case of blockage in the demodulation paths of a LoRaWAN gateway.

\begin{figure} [b!]
    \centering
    \includegraphics[trim=25 260 25 0, clip,width=1\columnwidth]{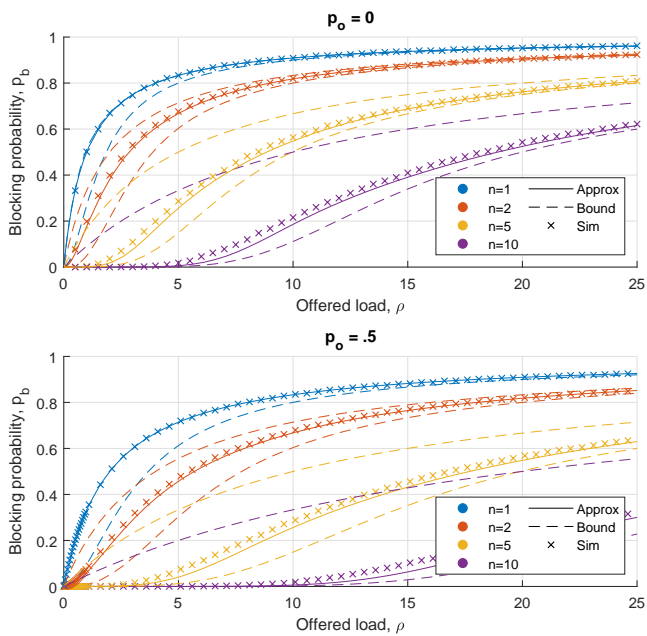}
    \caption{The blocking probabilities for the M/D/n/n~queue for expected offered load $\rho$ and intermittent arrival processes for $p_o=0$ and $p_o=0.5$. Notice that bounds hold and the approximation is close to the exact solution.%The non-blocking server utilization, $\zeta$ can be used to find the most cost-efficient number of servers, when the cost of blocking calls and not utilizing servers are equal.
    }
    \label{fig:Mres}
\end{figure}

\begin{figure} [tb]
    \centering
    \includegraphics[trim=25 0 25 0, clip,width=1\columnwidth]{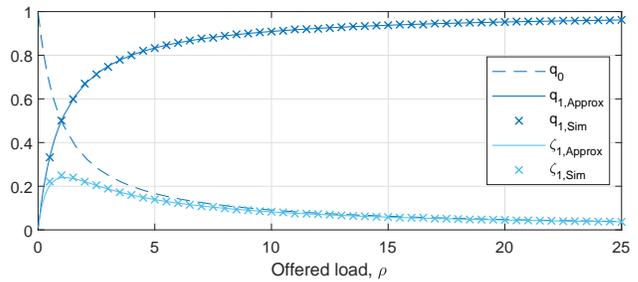}
    \caption{State probabilities, $q_0$ and $q_1$, server efficiency, $\eta=q_1$, and non-blocking server efficiency, $\zeta$, for the M/D/1/1~queue.
    }
    \label{fig:EtaMres}
\end{figure}

The approximated blocking probability and bounds for the M/D/n/n queue can be found in Fig. \ref{fig:Mres}. The approximation is close to the exact value, i.e. Erlang-B result. The server efficiency and non-blocking server efficiency are plotted in Fig. \ref{fig:EtaMres}. Here the result is exact owing to the timing analysis in Sec.~\ref{sec:timing} for $n=1$. 

\begin{figure} [tb]
    \centering
    \includegraphics[trim=25 260 25 0, clip,width=1\columnwidth]{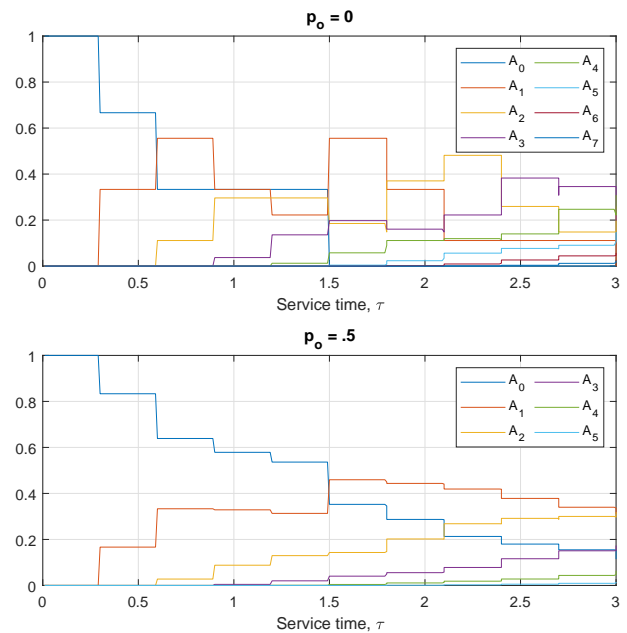}
    \caption{The probabilities $A_x$ of $x$ arrivals within the service time $\tau$ in D/D/n/n~queues for the configurations of the degenerate arrival process outlined in Tab.~\ref{tab:Dconfig}. $A_x$ does not depend on the number of servers $n$ so the results are equivalent for configurations ID~1$~\&~$3 and ID~2$~\&~$4, respectively.
    }
    % \includegraphics[trim=25 0 25 0, clip,width=1\columnwidth]{figures/Adegt.3.5.eps}
    % \caption{b) Pb for the different configurations versus tau.
    % }
    \label{fig:DA}
\end{figure}

\begin{figure} [tb]
    \centering
    \includegraphics[trim=25 25 25 0, clip,width=1\columnwidth]{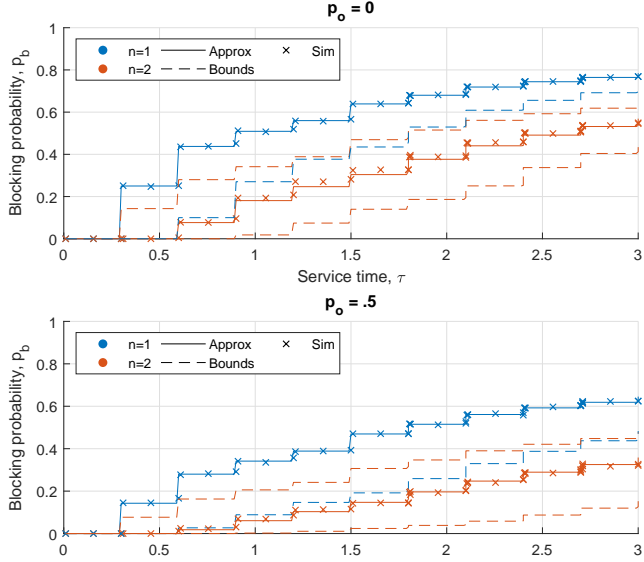}
    \caption{The blocking probability in D/D/n/n~queues as a step-wise function of the service time $\tau$ for the four configurations of the degenerate arrival process outlined in Tab.~\ref{tab:Dconfig}.}
    \label{fig:DPb}
\end{figure}

\begin{figure} [tb]
    \centering
    \includegraphics[trim=25 0 25 0, clip,width=1\columnwidth]{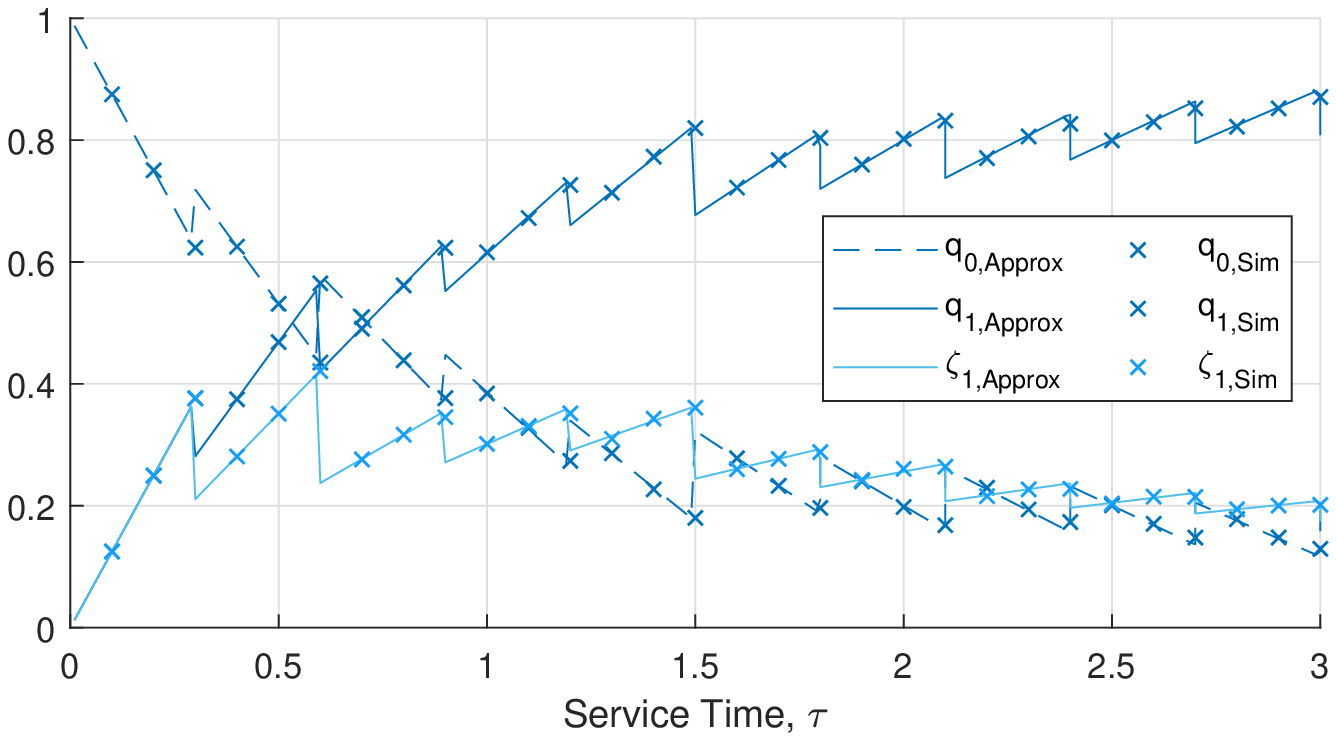}
    \caption{State probabilities, $q_0$ and $q_1$, server efficiency, $\eta=q_1$, and non-blocking server efficiency, $\zeta$, for the M/D/1/1~queue for ID~1 in Tab. \ref{tab:Dconfig}.
    }
    \label{fig:EtaDres}
\end{figure}

The arrival count as a function of the service time in a D/D/n/n queue changes as a nontrivial step-wise function of the service time $\tau$ and inter-arrival rate as depicted in Fig. \ref{fig:DA}. We observe that the index $x$ of the smallest neglible $A_x$ grows with the offered traffic load in Erlang. This also applies to the M/D/n/n queue, but in that case the count probability is a smooth function.
The blocking probability exhibits the same behaviour as depicted in Fig. \ref{fig:DPb}. The approximated blocking probability, here is also very close to simulated values. The Server efficiency and non-blocking server efficiency can be found in Fig. \ref{fig:EtaDres}. Notice that the mean arrival rate in the analysis of the D/D/n/n queue is fixed at $\lambda'~=~1/(0.3\cdot1/3+0.6\cdot1/3+1.5\cdot1/3~)=~1.25$.

{\rowcolors{1}{}{lightblue}
\begin{table}[tb]
\centering
\begin{tabularx}{.9\columnwidth}{|>{\raggedright\arraybackslash}X|>{\centering\arraybackslash}X>{\centering\arraybackslash}X>{\centering\arraybackslash}X>{\centering\arraybackslash}X>{\centering\arraybackslash}X>{\centering\arraybackslash}X>{\centering\arraybackslash}X>{\centering\arraybackslash}X|}
\hline
ID &n  & $t_1$  & $t_2$  & $t_3$  & $p_1$ & $p_2$  & $p_3$ &$p_o$\\ \hline
1 &1 &0.3  & 0.6  & 1.5 & 1/3 & 1/3 & 1/3 &0\\
2 &2 &0.3  & 0.6  & 1.5 & 1/3 & 1/3 & 1/3 &0\\
3 &1 &0.3  & 0.6  & 1.5 & 1/3 & 1/3 & 1/3 &0.5\\
4 &2 &0.3  & 0.6  & 1.5 & 1/3 & 1/3 & 1/3 &0.5\\ \hline
\end{tabularx}
\caption{Various degenerate arrival process configurations.}\label{tab:Dconfig}
\end{table}
}    

\begin{figure} [tb]
    \centering
    \includegraphics[trim=25 260 25 0, clip,width=1\columnwidth]{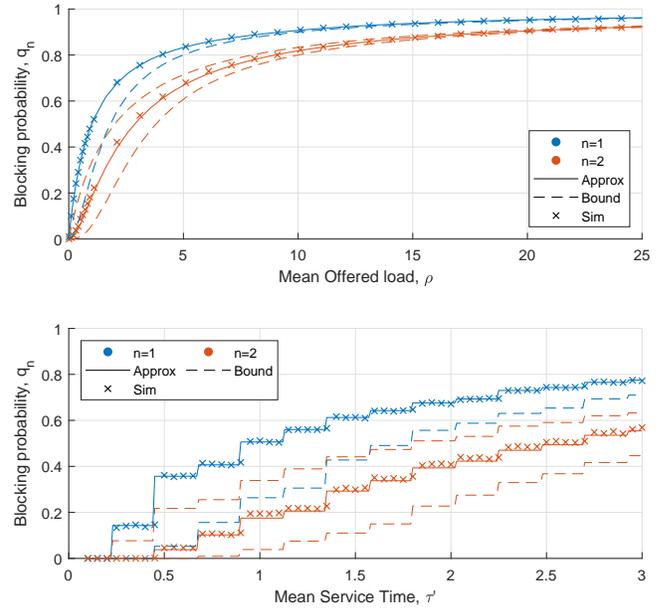}
    \caption{$P_b$ for Markovian (above) and degenerate (below) arrival processes and a service process that is defined by $p_x\in~\boldsymbol{p}=\{\dfrac{1}{2},~\dfrac{1}{2}\}$ and $\tau_x\in~\boldsymbol{\tau}=\{\dfrac{2}{3},~\dfrac{4}{3}\}$. The degenerate arrival processes in this example is ID~1$\&$ID~2 where $\boldsymbol{\tau}$ is scaled to achieve the mean service rate.
    }
    \label{fig:GGres}
\end{figure}

In Fig. \ref{fig:GGres} the blocking probabilities for the M/D/n/n queue and D/D/n/n queue are plotted for a service process that is defined by $p_x\in~\boldsymbol{p}=\{\dfrac{1}{2},~\dfrac{1}{2}\}$ and $\tau_x\in~\boldsymbol{\tau}=\{\dfrac{2}{3},~\dfrac{4}{3}\}$. The results are a close approximation, indicating the \eqref{eq:ggnn} holds.

The results of our investigation are straightforward to interpret as trade-offs in our motivating example of  the blocking probability of demodulation paths in a LoRa receiver; Clearly a larger number of demodulation paths $n$ yields a lower blocking probability, however the cost of increasing $n$ should yield an equivalent gain in performance. In general the distribution of the arrival-process is of course of dire importance. We observe a higher blocking probability $P_b = .25$ in the M/D/2/2 queue than $P_b = .068$ in the D/D/2/2 queue (configured as ID~2 in \ref{tab:Dconfig}) for $\tau=1$ and $\lambda' = .8$. It is also evident that given Degenerate arrival times tuning the service time relatively little can yield large gains.

\section{Conclusion}
In this paper, a framework for modelling G/D/n/n queues was presented and we looked at the case of blocking demodulation paths in LoRaWAN receivers.
%In essence, we may tune the blocking probabilities and efficiency of queues by tuning the mean service time in relation to the arrival rate. 
The proposed framework assesses the arrival count in service periods to model blocking probabilities in G/D/n/n quite accurately. Bounds valid for the blocking probability in G/D/n/n queues were also presented.
%which is a metric that takes into account both effects. 
In essence, we showed how blocking probability depends on the arrival process, which is therefore essential for describing the exact blocking probability for e.g. demodulation in LoRaWAN receivers. In general, increasing $n$, decreases the blocking probability as one would expect intuitively. Exponential inter-arrival times are often assumed for the arrival process of communication networks, but for example in the case of the LoRaWAN receiver, this transmission process will be filtered by capture effect, yielding a non-Markovian distribution at the demodulation paths.

The framework depends on finding counting functions, which may be relatively easy to find by induction given the tools available today for integrating symbolic expressions. Moreover, the framework is directly applicable when the distribution of the inter-arrival times is given numerically as 'binned' or Degenerate approximation of the arrival process, but it is not available on an explicit analytical form. The methodology was shown to yield close approximates of the well known results for the M/G/n/n queue along with close approximates for the D/D/n/n queue. 
The framework was extended to G/G/n/n-queues, but verification was limited to a small set of service times due to simulation complexity.
%A specific example of demodulation in LoRa transceivers was given as motivation for the model, but it is widely applicable.
% yielding some expressions for finding the loss and steady-state probabilities of the G/G/n/n queue.

% it was shown that steady state analysis of the M/D/n/n queue yields equivalent results to the M/M/n/n queue, ie. the Erlang B queue, and that this result could be generalised to the M/G/n/n queue. Furthermore, 

%Furthermore, the applicability of the results in advanced queuing scenarios common in engineering problems was analysed and easy-to-use expressions yielded.

%An intuitive reasoning for the loss being equivalent is that the average service rate of the servers is the same for the two systems and there is no buffer of queued arrivals to be served when the exponentially distributed service times deviates from the mean. Note, that the Engset formula is not applicable to M/D/n/n+N queues.
%The above intuition does not suffice to explain the equivalent steady-state probability distributions. 

% \section*{Acknowledgment}
% The authors would like to thank... grants

\bibliographystyle{IEEEtran}
\bibliography{bibtex}

\end{document}